
\input harvmac.tex
\font\Bigit=cmti10 scaled \magstep2
\noblackbox\parindent=0pt
\def\a{\alpha}\def\b{\beta}\def\g{\gamma}
\def\d{\delta}\def\D{\Delta}\def\e{\epsilon}
\def\z{\zeta}\def\t{\theta}

\def\l{\lambda}\def\o{\omega}
\def\O{\Omega}\def\s{\sigma}
\def\vphi{\varphi}
\def\pa{\partial}\def\na{\nabla}
\def\phidot{{\dot\phi}}
\font\large=cmr10 scaled\magstep 1
 2
 3
 3
\rightline {UPR-523T}

\bigskip
\centerline{{\large NON-PERTURBATIVE EFFECTS IN 2-D STRING THEORY }}
\centerline{{\large OR}}
\centerline{{\large BEYOND THE LIOUVILLE WALL }}
\vskip.3in
\nopagenumbers
\centerline {RAM BRUSTEIN and BURT A. OVRUT}
\bigskip\baselineskip=10pt
\centerline{Department of Physics}
\centerline{University of Pennsylvania}
\centerline{Philadelphia, PA 19104}
%


\vskip.8in

\baselineskip=12pt plus 2pt minus 1 pt
 {\centerline{ABSTRACT}
\bigskip
\noindent
We discuss continuous and discrete sectors in the collective field theory
of $d=1$ matrix models. A  canonical Lorentz invariant field theory
extension of  collective field theory is presented and its classical
solutions in Euclidean and Minkowski space are found. We show that
the discrete, low density,
sector of collective field theory includes single eigenvalue Euclidean
instantons which tunnel between different vacua of the extended theory.
We further show that these ``stringy" instantons induce  non-perturbative
effective operators of strength  $e^{-{1\over g}}$ in the extended theory.
The relationship of the world sheet description of string theory and
Liouville theory to the effective space-time theory is explained. We
also comment on the role of the discrete, low density,
 sector of collective field theory
in that framework.}
\bigskip\bigskip

\footnote{}{{\sevenrm email: ramy@mohlsun.physics.upenn.edu}}
 \footnote{}{{\sevenrm email: ovrut@penndrls.upenn.edu}}

\vfill\eject
\footline={\hss\tenrm\folio\hss}
\baselineskip=16pt plus 2pt minus 1pt

\newsec{INTRODUCTION}

Non-perturbative aspects of string theory are an essential piece
of information needed to make the comparison
between string theory predictions and the real observable world.

Matrix models, and especially $d=1$ matrix models
\ref\grossa{ D. J. Gross, N. Miljkovic,  Phys.Lett.B238 (1990) 217\semi
 P. Ginsparg and J. Zinn-Justin, Phys. Lett. B240 (1990) 333\semi
 E. Brezin, V. Kazakov and Al. B. Zamolodchikov, Nucl. Phys. B338 (1990) 673.},
 offer  a unique opportunity to obtain some  insight into non-perturbative
string theory. Certain matrix models have associated with them very simple
string theories with a low number of degrees of freedom, propagating in a low
number of space-time dimensions.

The $d=1$ matrix model is the most complicated  matrix model
which  can be  solved exactly.
On the other hand, it describes the simplest space-time dynamics which
is still interesting. In the double scaling limit, the $d=1$ matrix model
 describes strings propagating in one time dimension
and one spatial dimension. An equivalent description is given in terms
of a bosonic collective field theory in $1+1$ dimensions of one massless field
\ref\das{S.R. Das and A. Jevicki, Mod. Phys. Lett. A5 (1990) 1639.},
\ref\joeb{ J. Polchinski,  Nucl.Phys. B346 (1990) 253.}. Notable features
of collective field theory is that  the kinetic energy is not canonical
and it is not Lorentz invariant. Yet another
equivalent description is in terms of $1+1$ dimensional fermionic field
theory
\ref\igor{D. J. Gross and I.R. Klebanov,
Nucl.Phys. B352 (1991) 671.}--\nref\wadsen{ M. Sengupta and  S. R. Wadia,
Int.J.Mod.Phys. A6 (1991) 1961.}\ref\wadia{A. Dhar, G. Mandal and S. R. Wadia,
IAS Princeton preprint, IASSNS-HEP-91-89 (1992).}.
This description is useful to obtain the general set of classical
solutions of the field theory
\ref\joe{J. Polchinski,  Nucl.Phys. B362 (1991) 125. }.
In the bosonic theory, scattering amplitudes were calculated in perturbation
theory
\ref\antalb{K. Demeterfi, A. Jevicki and J. P. Rodrigues,
Nucl.Phys. B362 (1991) 173\semi
K. Demeterfi, A. Jevicki and J P. Rodrigues, Nucl.Phys. B365 (1991) 499.},
\ref\antalc{K. Demeterfi, A. Jevicki and J. P. Rodrigues,
Mod. Phys. Lett. A6 (1991) 3199.}.
Similar calculations were done in the matrix model formulation and in
the fermionic field theory in refs.\igor,
\ref\grig{D. J. Gross and I. R.  Klebanov,
Nucl. Phys. B359 (1991) 3.}, and ref.
\ref\huld{ G. Moore,  Nucl.Phys. B368 (1992) 557.}.
In this paper, we use the bosonic collective  field theory because it has
a more  transparent  space-time description.

As stated above, the $d=1$ matrix models, or the equivalent field theories
have the power to describe non-perturbative phenomena in the associated
$1+1$ string theories. This is interesting by itself.
However, there may well be general  features of non-perturbative string
theory that are common to all string theories, including
more complicated theories in higher dimensions such as $d=4$.
By studying the generic features of non-perturbative
behaviour in $1+1$ dimensional string theories,
one may learn about more realistic 4-dimensional string theories.
It is of interest to ask whether or not there are any indications in string
theory of relevant, non-perturbative behaviour. The
answer \ref\shenker{S. Shenker, Cargese Workshop on Random
Surfaces, Quantum Gravity and Strings, Cargese, France (1990).}
is yes!

First recall that in quantum field theory there is a well known connection
 between the  large order behaviour of amplitudes and non-perturbative effects.
Typically, amplitudes grow as   $G!$  where $G$ is the number of loops,
while non-perturbative effects have strength
$e^{-{1\over g^2}}$, where $g$ is the coupling parameter of the
theory. Both of these facts follow from
 the existence of  non-trivial
classical solutions  of the equations of motion of the field theory
(or related field theory)  in Euclidean space, i.e. instantons.
The magnitude of the non-perturbative
effects due to non trivial solutions in a field theory with one dimensionless
coupling parameter $g$ can be estimated using a simple scaling argument.
Since the coupling parameter  in this case  can be scaled away, the
action can be written as $S(\phi,g)={1\over g^2} \widetilde S(\widetilde\phi)$,
where $\widetilde S$ does not depend on $g$.
Therefore, any classical Euclidean solution with finite action has an action
of order ${1\over g^2}$.
The magnitude of large order terms
in the perturbative expansion can also be estimated by counting Feynman
diagrams. The number $G!$ basically comes from the number of diagrams.

Large  order growth of perturbative
amplitudes is a common feature of matrix models
 and more complicated string theories\shenker.
For a review of large order behaviour of matrix model amplitudes
see ref.\ref\zinn{ P. Ginsparg and
J. Zinn-Justin, Phys. Lett. B255 (1991) 189.}.
All matrix models, as well as the critical bosonic string theory in 26
dimensions, exhibit a strange phenomenon. The magnitude of $G$'th order
amplitudes in  perturbation theory  grow like $(2G)!$. We discovered that this
simple,  unalarming fact has far reaching consequences, which we explain in
this paper.

It turns out that, in much the same way  as $G!$ behaviour corresponds to
$e^{-{1\over g^2}}$ non perturbative effects in quantum field theory,
in matrix models the large order  $(2G)!$ behaviour
would correspond to non-perturbative effects of strength $e^{-{1\over g}}$.
How do these peculiar effects arise?
In matrix models, there is a new type of instanton, involving
a single eigenvalue, that is responsible for these effects.
For a discussion of one eigenvalue instantons see \shenker,\zinn.
They were also discussed in
\ref\lecht{O. Lechtenfeld, Int. J. Mod.Phys. A7 (1992).} and in the
context of supersymmetric matrix models in refs.
\ref\dabh{A. Dabholkar, Nucl.Phys. B368 (1992) 293.} and
\ref\hans{ J. D. Cohn and  H. Dykstra, Mod.Phys.Lett. A7 (1992) 1163.}.

In view of the above scaling argument in quantum field theory,
it is of interest to ask how an action of order ${1\over g}$ can ever arise.
The answer is that, in matrix models, the associated effective action
does not obey the same scaling argument,
$S(\phi,g) \ne {1\over g^2} \widetilde S(\widetilde\phi)$. Instead,
one finds that $g$ cannot be completely scaled out of $\widetilde S$
due to ``scale breaking terms". That is
$S(\phi,g) = {1\over g^2} \widetilde S(\widetilde\phi,g)$. It follows
that a non-trivial solution can be a function of $g$. Furthermore, if
for such a solution $\widetilde S\sim g$, then $S\sim {1\over g}$.
This is exactly what happens for one eigenvalue instantons.

The questions that we set out  answer in this paper are:

1. Is there a canonical, Lorentz invariant effective field theory that
describes matrix models and collective field theory?

2. What is the relation of single eigenvalue instantons to such a field theory,
and in what sense do they describe a tunneling phenomenon?

3. Do these instantons induce calculable non-perturbative operators in the
effective field theory?

4. What is the relationship of all of the above to string theory?

The main line of this paper is built around the answers to these questions.
In  section 2 we discuss various facts about matrix models and collective
field theory. In section 3, the double scaling limit is presented
in the context of collective field theory. We then discuss the high density
limit of this theory and the static solution of its equations of motion.
Similarly, we define the low density , finite eigenvalue limit.
We show that, in Euclidean space, there is a single eigenvalue instanton
solution of the equations of motion which has action ${\pi\over g}$.

Section 4 is devoted to extending the collective field theory, which is a
non-canonical, non-Lorentz invariant theory of a single field $\phi$,
to a canonical, Lorentz invariant effective field theory of two
fields $\z$ and $D$. This theory exhibits the scale breaking terms
responsible for the unusual action, $S\sim {1\over g}$, of the instantons.
This answers question 1. above in the affirmative.
In section 5 we discuss, in detail, the vacuum solutions of the
effective field theory and how these include wormhole-like configurations.
These configurations look like two identical Liouville vacua linked by
a single eigenvalue instanton. The normalized action of these
configurations is $S={\pi\over g}$. Thus question 2. is answered.
We then show that these configurations
induce calculable operators in the $\z$, $D$ effective theory with strength
$e^{-{\pi\over g}}$, as conjectured by Shenker. This answers question 3.
Until this point,
we have been discussing matrix models only. In section 6 we review
the relationship of these double scaled matrix models to $1+1$ dimensional
string theories, and argue that these string theories must include the
new vacuum configurations, single eigenvalue instantons, and induced
operators of strength $e^{-{\pi\over g}}$.  This answers question 4.
and concludes the main content of the paper. Finally, in section 7 we
present some possible directions in which our results can be extended.

This paper is an extended and detailed version of
\ref\il{R. Brustein and B. Ovrut, Penn. preprint, UPR-522T (1992).}.
\newsec {MATRIX MODEL AND COLLECTIVE FIELD THEORY}

In this section we review a few well known facts and a few less well known
facts  about the $d=1$ matrix model and collective field theory and
present them in a form appropriate for the following  discussion of
string theory.

\subsec{Matrix Model}

The fundamental variables of a time dependent, hermitian matrix model are
$N\times N$ hermitian matrices $M(t)$. Their dynamics is described by the
Lagrangian
\eqn\matlag{ L({\dot M},M)=\half  Tr{\dot M}^2 - V(M)}
where, in general, $V$ is a finite polynomial
\eqn\matpot{V(M)=\sum_n g_n Tr M^n }
and $g_n$ are real coupling parameters. Clearly the mass dimension of
$M$ is $-\half$ and, hence, the couplings $g_n$ have positive mass dimensions.
The conjugate momenta to $M$ are the $N\times N$ hermitian matrices
$\Pi_M(t)={\dot M}$. It follows that the associated Hamiltonian is given by
\eqn\matham{ H(\Pi_M,M)=\half Tr\Pi_M^2+V(M) }
The partition function of the hermitian matrix model can be written in
terms  of a path integral
\eqn\hmatpart{Z_N(g_n)=\int [dM][d\Pi_M] \
e^{\ i\int dt \left\{\Pi_M {\dot M} - H(\Pi_M,M)\right\} } }
The $\Pi_M$ integral is Gaussian and can be done explicitly. The result is
\eqn\matpart{Z_N(g_n)=\int dM\ e^{\ i\int dt L({\dot M},M)} }
One now notes that matrices $M$ remain hermitian under the transformation
$M\rightarrow U M U^\dagger$. The Lagrangian is invariant under such
transposition as long as $U\in U(N)$. Therefore, as far as the partition
function is concerned, the matrices $M$ can always be expressed in terms
of their $N$ real eigenvalues $\l_i$. Furthermore, only correlation functions
of operators that are $U(N)$ singlets are considered, since it is  these
singlet operators that correspond to string theory. It follows that the
entire singlet sector theory can be completely expressed in terms of the
eigenvalues $\l_i$. We do this explicitly in the next section.
Another, equivalent, formulation of the same theory is the collective
field representation which we discuss in section (2.3).

\subsec{Effective Theory for Eigenvalues}

We proceed to evaluate
the partition function \matpart\ in terms of the eigenvalues
$\l_i(t), i=1,\dots,N $ of the matrix. The change of variables from
$M$ to its eigenvalues and angular variables is non-linear. It is, therefore,
difficult to proceed directly from the partition function \matpart. It is
simpler to return to expression \hmatpart\ and to use the formalism developed
in ref.\ref\sakita{B. Sakita, Quantum Theory of Many-Variable Systems and
Fields, World Scientific, Singapore 1985. }.
The result is that, up to an unimportant normalization
factor coming from the integration over angular variables
\eqn\eigpart{Z_N(g_n)=\int[d\l_i][d\Pi_{\l_i}]
e^{i\int dt\Biggl\{
\sum\limits_i \Pi_{\l_i} {\dot\l_i} -H_{eff}({\Pi_{\l_i},\l_i})
\Biggr\} } }
where $H_{eff}$ can be determined as follows. The effective
Hamiltonian operator in the $\l_i$ representation is found to be
\eqn\sakitaa{{\hat H}_{eff}={\cal J}^{\half}\ {\hat H}\ {\cal J}^{-\half} }
where ${\cal J}$ is the Jacobian for the change of variables from the
matrix variables to the eigenvalue variables
\eqn\vander{{\cal J}=\prod\limits_{i<j}(\l_i-\l_j)^2}
and
\eqn\hamil{{\hat H}=\sum\limits_i -\half {\pa^2\over \pa\l_i^2 }+V(\l_i) }
Here $V(\l_i)$ is the potential \matpot\ written in terms of the eigenvalues.

The effective Hamiltonian should be hermitian in the new variables so
care should be exercised in interpreting Eq.\sakitaa.
The Hamiltonian  ${\hat H}_{eff}$ can  be evaluated explicitly
\eqn\hamilt{{\hat H}_{eff}=
\sum\limits_i \Biggl[-\half {\pa^2\over \pa_{\l_i^2} }+
\half\left( \sum\limits_{j\ne i} {1\over \l_i-\l_j}\right)^2
+V(\l_i)\Biggl] }
Noting that in the $\l_i$  representation the operator
$\Pi_{\l_i}=-i {\pa\over \pa\l_i}$, it follows that the associated
classical Hamiltonian in Eq.\eigpart\ is given by
\eqn\hamilto{H_{eff}=\sum\limits_i \Biggl[\half\Pi^2_{\l_i} +
\half\left( \sum\limits_{j\ne i} {1\over \l_i-\l_j}\right)^2
+V(\l_i) \Biggr]}

Inserting this expression into \eigpart\ and performing the Gaussian
$\Pi_{\l_i}$ integration, we find that

\eqn\eigpap{Z_N(g_n)=\int[d\l_i]
e^{i\int dt L_{\rm eff}({\dot \l}_i,\l_i)} }
where
\eqn\eiglageff{L_{\rm eff} ({\dot \l}_i,\l_i)=
\half \sum\limits_i{\dot\l_i}^2
-V_{{\rm eff}}(\l_i)}
and
\eqn\veff{V_{{\rm eff}}=V_{{\rm coll}}(\l_i)+V(\l_i) }
The induced term in the effective potential, $V_{{\rm coll}}$, is
\eqn\vcollec{V_{{\rm coll}}(\l_i)=
\half \sum\limits_i
\left(\sum\limits_{j\ne i} {1\over (\l_i-\l_j)}\right)^2}
whereas the original potential in terms of the eigenvalues, $V(\l_i)$ is
\eqn\orpot{V(\l_i)=\sum\limits_i\sum\limits_n g_n \l_i^n }

The classical equations of motion of the theory are then given by
\eqn\motmat{ {d^2\l_i\over dt^2} =
 -{d\over d\l_i}V(\l_i)- {d\over d\l_i} V_{{\rm coll}}(\l_i) \ \ \ \
i=1,\dots,N}
In general the solutions of Equations \motmat\ are very complicated, but
there are some conditions on the density of eigenvalues that
make them tractable.

We reiterate, for emphasis,  that  Eqs.\motmat\  are the complete and unique
equations  of motion of the matrix model, restricted to the singlet sector.
They have to be satisfied in any other representation of the theory.

\subsec{Collective Field Theory}

A useful step on the road from  the matrix model to string theory is
 collective field theory \das. Relevant references are
\ref\sakicki{A. Jevicki and B. Sakita,
Nucl. Phys.  B165 (1980) 511.} --\nref\cohn{J. D. Cohn and S.P. De Alwis,
Nucl.Phys.B368 (1992) 79.}
\ref\karb{D. Karabali and B. Sakita,  Int. J. Mod. Phys. A6 (1991) 5079.} and
\ref\antala{ A. Jevicki, Nucl.Phys. B376 (1992) 75.}.
The idea is to  start from the matrix model and, by
performing a series of changes of variables, arrive at a field theory
representation of the matrix model. We review here the derivation of
the collective field theory Lagrangian.

We restrict the range of the eigenvalues $\l_i$  to be finite and
impose periodic boundary conditions. That is
\eqn\rngeig{ -\half L\le \l_i \le \half L}
A continuous spatial variable, $x$, is now introduced which also satisfies
\eqn\rngx{ -\half L\le x \le \half L}

One can now define the eigenvalue density, $\phi$, by
\eqn\denn{\phi(x,t)={1\over N} \sum\limits_{i=1}^N\delta\left(x-\l_i(t)\right)}
Note that $\phi$ satisfies the constraint
\eqn\xx{\int\limits_{-\half L}^{\half L} dx \phi(x,t)=1}
The eigenvalue density is  also called the collective field.

Since the number of degrees of freedom of the system is $N$,
not all $\phi(x,t)$ are independent. The $N$ independent variables associated
with $\phi$ are the Fourier components
\eqn\colvar{\phi_{k_n}(t)=\int dx e^{-ik_nx}\phi(x,t)}
where
\eqn\remo{k_n ={2 \pi n \over L}\ \ \ \
n=\pm  1,\pm 2\dots,\pm { N\over 2}  }
The highest momentum is $ k_{max} = {\pi N\over L} $.

We proceed to evaluate the partition function \matpart\  in terms of the
collective field $\phi$. It is easier to return to expression \hmatpart\ and
to use the formalism developed in ref.\sakita. The partition function
is given by
\eqn\hmatcoll{Z_N(g_n)=\int [d\phi][d\Pi_\phi] e^{i\int dt dx \left\{
\Pi_\phi {\dot\phi}-{\cal H}_{eff}(\Pi_\phi,\phi)\right\} } }
where $\Pi_\phi$ is the conjugate momentum of the field $\phi$
and ${\cal H}_{eff}$
can be determined as follows. Let ${\cal J}[\phi]$ be the Jacobian associated
with the change of variables from $\l_i$ to $\phi$.
Since ${\cal J}$ can be quite complicated to evaluate directly,
it is more convenient to  express it in terms of  two functions
$\O(x,y,\phi),\o(x,\phi)$ defined as follows.
\eqn\Womega{\O(x,y,\phi)=\sum_{i,j}{\delta \phi(x)\over \delta\l_i}
{\delta \phi(y)\over \delta\l_j}
={1\over N}{\pa\over\pa x}{\pa\over\pa y} [\d(x-y)\phi(x)]}
\eqn\womega{
\o(x,\phi)=-\sum_i{\delta^2\phi(x)\over \delta\l_i^2 }=
2N{\pa\over\pa x}[\phi(x)\ {\bf P}\int dy {\phi(y)\over x-y}] }
where ${\bf P}$ stands for the principal part.
The Jacobian can be shown to satisfy
\eqn\JJ{\int dy \O(x,y,\phi) {\delta\over \delta\phi(y)} \ln{\cal J}=
{\delta\O\over\delta\phi}+\o }
This equation can be solved for ${\cal J}$. The result is
\eqn\newjackcity{\ln[{\cal J}]=
 N^2 \int dx dy\ \phi(x)\ \ln|x\!-\!y\!-\!i\e|\ \phi(y) }
where $\e$ is a regulator that is taken to zero eventually.

The Hamiltonian density of the theory is then given by
\eqn\chicken{ {\cal H}_{eff} =\half\ \Pi_\phi\ \O\ \Pi_\phi+
{1\over 8}\ {\d \ln [{\cal J}]\over \d \phi}\ \O\
{\d \ln[{\cal J}]\over \d \phi}
 +N(V(x)-\mu_F) \phi }
where $\mu_F$ is a Lagrange multiplier introduced
to enforce the constraint \xx\ and $V$ is the potential obtained from the
eigenvalue potential \orpot\ expressed in the new variables,
$V(x)=\sum\limits_n g_n x^n$.

Using Eqs.\Womega,\newjackcity, and the identity
\eqn\idnsak{
\half \int dx \phi(x,t)\ {\bf P}\int dy  {\phi(y,t)\over x-y}\ {\bf P}\int dz
{\phi(z,t)\over x-z}= {\pi^2\over 6}\int dx \phi^3(x,t) }

This Hamiltonian density becomes
\eqn\ham{\eqalign{   {\cal H}_{eff} =
& {1\over 2 N} \pa_x \Pi_\phi\ \phi\ \pa_x\Pi_\phi
+{N^3 \pi^2\over 6} \biggl[\phi^3
-\phi(x)\phi(y)\phi(z)\d(x-y)\d(x-z)\biggr]
\cr + & N(V(x)-\mu_F)\phi   \cr } }

The significance of the peculiar term
$ N^3\phi(x)\phi(y)\phi(z)\d(x-y)\d(x-z)$  can be understood
by evaluating it using
$\phi(x,t)= {1\over N} \sum\limits_i \d\left( x-\l_i(t)\right)$.
The result is
\eqn\cntra{ \int dx \phi(x)\phi(y)\phi(z)\d(x-y)\d(x-z)=  N\d^2(0) }

This term then can be thought of as classical counter term consistent with the
underlying  matrix model.
It  ensures that  field configurations  of the form
 $\phi={1\over N}\sum\limits_i\d\left(x-\l_i(t)\right)$
have finite energy by subtracting
classical self-energy contributions. It's effect on smooth field
configurations is negligible. We will stop writing this term down at this
point. It is however important in discussing the low density limit
in the next section.

It is useful to rescale the various quantities
according to
\eqn\sclrule{\eqalign{&  \phi\rightarrow {1\over\sqrt{N}}\phi,\ \ \
 \Pi_\phi\rightarrow {1\over N} \Pi_\phi,\ \ \
 x \rightarrow \sqrt{N} x \cr &
V\rightarrow N V,\ \ \ \mu_F\rightarrow  N \mu_F \cr }}
The Hamiltonian  density then becomes
\eqn\bacon{{\cal H}_{eff} =
  {1\over 2 N^2}\pa_x\Pi_\phi\ \phi\ \pa_x\Pi_\phi
+{N^2 \pi^2\over 6} \phi^3+N^2(V(x)-\mu_F)\phi  }
Inserting this expression into Eq.\hmatcoll\ and noting that the first term in
the Hamiltonian can be written as
\eqn\firstterm{\int dx {1\over 2 N^2}\pa_x\Pi_\phi\ \phi\ \pa_x\Pi_\phi
=\int dx dy \Pi_\phi(x,t)\ {1\over 2 N} \Omega(x,y,\phi)\ \Pi_\phi(y,t) }
we can perform the Gaussian $\Pi_\phi$ integration. The result is
\eqn\partcoll{ Z_N(g_n) =
  \int [d\phi]\ {\rm det}\O^{-\half}
e^{i  N^2\int dt  L_{\rm eff}
({\dot\phi},\phi) }   }
where
\eqn\lag{L_{\rm eff}= \int dx\left\{
\half{\int\limits^x\phidot \int\limits^x\phidot \over \phi}
-{\pi^2\over 6} \phi^3- (V(x)-\mu_F)\phi \right\} }
Here the contact terms in Eq.\cntra\ are omitted.
Note the appearance of the factor ${\rm det}\O^{-\half}$ which comes from
doing the  Gaussian integral over the conjugate momentum $\Pi_\phi$.

A good check on the validity of Eq.\partcoll\ is to substitute the rescaled
version of Eq.\denn\
$\phi=\sum\limits_i\d\left(x-\l_i(t)\right)$
into it. The result should be identical to Eq.\eigpap. Inserting the above
rescaled version of \denn\  into \lag\
(including the contact terms) and using
\eqn\daretocompare{
\half \int dx \phi(x)\ {\bf P}\int dy{\phi(y)\over x-y}\ {\bf P}\int dz
{\phi(z)\over x-z}=
\half \sum\limits_i\left(\sum\limits_{j\ne i} {1\over (\l_i-\l_j)} \right)^2}
we find that $L_{\rm eff}({\dot\phi},\phi)$ is equal to $
L_{\rm eff}({\dot\l}_i,\l_i)$
in \eiglageff. Similarly, one can show that
$\int [d\phi]{\rm det}\O^{-\half}=\int[d\l_i]$ and, hence the two partition
functions are equal, as they must be. We note for future use that one can
 use the identity \idnsak\  in the other direction and deduce
that an equivalent form of the collective potential
in Eq.\vcollec\ is
\eqn\newvcoll{ V_{{\rm coll}}(\l_i)=
{\pi^2\over 6} \sum\limits_i\left(\sum\limits_{j\ne i} \d(\l_i-\l_j)\right)^2}

At this point, however, we must be careful. Recall that the eigenvalues $\l_i$
are independent variables, whose equation of motion follow from
varying the action in Eq.\eiglageff\ with respect to each of the $\l_i's$.
However, as discussed above, for finite $N$ the collective field
$\phi(x,t)$ is highly constrained. It does  not correspond to an infinite
number of degrees of freedom. Hence, the $\phi$  equation of motion is
 not obtained by varying the action with respect
to $\phi$. The correct procedure is the following.
Start with the equations of motion for the eigenvalues $\l_j$
\eqn\motev{ {\d\over \d\l_j(t)} S_{\rm eff}[ {\dot\l}_i,\l_i]=0 }
where $S_{\rm eff}[ {\dot\l}_i,\l_i]=\int dt L_{\rm eff}( {\dot\l}_i,\l_i)$.
Then by using the rescaled version of \denn\ and the fact that
$L_{\rm eff}( {\dot\phi},\phi)=L_{\rm eff}( {\dot\l}_i,\l_i)$ convert \motev\
to an equation for $\phi$. The result is
\eqn\motphi{ {\pa\over\pa y} {\d S_{\rm eff}[ {\dot\phi},\phi]\over
\d\phi(y,t) }{}_{|_{y=\l_j(t)}}=0 \ \ \ \ j=1,...,N }

Note that there are $N$ equations of motion since $y$ must be evaluated
at $\l_j$ for all $j=1,...,N$. Furthermore, note that $\phi's$ satisfying the
naive $\phi$ equation of motion
\eqn\naive{{\d S_{\rm eff}[ {\dot\phi},\phi]\over
\d\phi(y,t) } =0 }
also satisfy  Eq.\motphi. However there are solutions
of Eq.\motphi\ which do not satisfy Eq.\naive. We return to this important
point later.

\newsec{LARGE $N$ LIMIT }

In the previous section, we discussed matrix models, in both the $\l_i$
and $\phi$ representations for finite $N$. Furthermore,
the $\l_i$'s and $x$ were restricted to satisfy $-\half L \le\l_i\le \half L$
and $-\half L \le x \le \half L$ respectively for finite $L$. In this section
we discuss the limits $N\rightarrow\infty$ and $L\rightarrow\infty$. We find
it convenient to take the limit $L\rightarrow\infty$ at the outset. Henceforth,
$L\rightarrow\infty$ and $\l_i$ and $x$ satisfy $-\infty \le\l_i\le \infty$
 and $-\infty \le x \le \infty$. We now want to let $N\rightarrow\infty$.
However, as we  proceed to show,
there are several inequivalent ways in which the large $N$ limit can be taken.

\subsec{Double scaling limit}

Before taking the $N\rightarrow\infty$ limit, it is essential to specify
the dependence on $N$ of the coupling parameters $g_n$ in the potential
\matpot. As usual, if we do not choose the $N$ dependence of the couplings
in any special way, the resulting large $N$ limit is a free theory
\ref\rgmtx{R. Brustein and S. De Alwis,
Proceedings of the  PASCOS-91 Int. Symp., Boston 1991,
P. Nath and S. Reucroft Ed., World Scientific.},
\ref\heh{E. Brezin and J. Zinn-Justin, Ecole Normale preprint,
LPTENS-92-19, 1992.}. The choice of $N$ dependence of couplings (as
$N\rightarrow\infty$) that turns out to be most relevant for string theory
is called the double scaling limit. It involves specifying the exact $N$
dependence of one coupling parameter. Which of the coupling parameters is
specified is not important.
The $\l_i$ representation in subsection  {\it (2.1)}
and  the $\phi$ representation  in subsection  {\it (2.2)} are equivalent.
Therefore the double scaling limit can be taken in either representation.
We find it more convenient to define and take this limit in the $\phi$
representation. Therefore, here and  throughout the rest of the paper,
we use the collective field representation.

To enable us to take the double scaling limit the potential, $V(x)$, has to
have a local maximum at some point $x^*$. For every value of $x^*$ one gets the
same double scaling limit, so the position of the maximum is unimportant.
Therefore, for simplicity, we set $x^*=0$. Then
$V(x)=V(0)-\half x^2+\cdots$, where, without loss of generality,
 we have chosen the coefficient of the second term to be $\half$. There is
a region $|x|\le x_{\rm max}$ where the higher order terms in $V(x)$
can be neglected. We restrict our attention to that region.
 Therefore, in that region $V(x)=V(0)-\half x^2$.
Inserting this expression into the Lagrangian \lag,
the combination $V(0)-\mu_F$ appears.
We denote $V(0)-\mu_F$ by $\half\mu$ and  assume that $\mu>0$.

The double scaling limit is defined by specifying the $N$ dependence of $\mu$
\eqn\dsl{N\mu={1\over g} }
so that $g$ remains finite as $N\rightarrow \infty$. The parameter $g$ is
related to the string coupling parameter, as we discuss later on. (For this
reason this parameter is often denoted $g_{\rm st} $ in the literature.).

It is convenient  at this point to again rescale
\eqn\ndsclrule{ \phi\rightarrow {1\over\sqrt{N}}\phi,\ \ \ \ \
 x  \rightarrow \sqrt{N}x }
This rescaling removes the  factor $N^2$ in Eq.\partcoll. Now take
the $N\rightarrow \infty$ limit. Then
$x=\sqrt{\mu}\rightarrow x= \sqrt{{1\over g}}$ and
the region $|x|\le x_{\rm max}$
gets blown up to the whole real axis
 $-\infty\le x \le \infty$.

The Lagrangian  density  in Eq.\lag\  now becomes
\eqn\lager{{\cal L}_{\rm eff}=
\half{\int\limits^x\phidot \int\limits^x\phidot \over\phi}
-{\pi^2\over 6} \phi^3-\half ({1\over g}-x^2)\phi }
The classical equations of motion derived using Eqs.\motphi\ and \lager\
are
\eqn\eqmotcoll{ {\pa\over \pa x} \left(
\int\limits^x dy \pa_t{\int\limits^y\phidot\over\phi}
-{1\over 2}{\int\limits^x\phidot \int\limits^x\phidot \over\phi^2}
 -{\pi^2\over 2} \phi^2 -\half ({1\over g}-x^2)\right){}_{|x=\l_i(t)} =0 }
where the index $i$ now runs over $i=,1,...,N\rightarrow\infty$.

At this point we realize  that to make sense of the equations of motion
within the double scaling limit, it is necessary to specify more accurately
the density structure of the eigenvalues $\l_i$.
The additional information lies in the
nature of the limit of $N\over L$ as
$L\rightarrow\infty$, $N\rightarrow\infty$, which has not been specified yet.

\subsec{High Density Limit}

The high density (HD) limit of collective field theory is defined as follows.
Consider  a region of  $x$, denoted $I$,  of length
$l(I)$. The number of
eigenvalues in this region is $N(I)$.
Then take the the double scaling limit in such a way that
\eqn\hdl{ {N(I)\over l(I)}\rightarrow\infty}
It is clear that in this region of $x$ there are an infinite number of
eigenvalues. The collective field $\phi$ now has an infinite number of degrees
of freedom  and the classical equations of motion \eqmotcoll\ become simply
\eqn\eqmotCOLL{ {\pa\over \pa x} \left(
\int\limits^x dy \pa_t{\int\limits^y\phidot\over\phi}
-{1\over 2}{\int\limits^x\phidot \int\limits^x\phidot \over\phi^2}
 -{\pi^2\over 2} \phi^2 -\half ({1\over g}-x^2)\right) =0 }

The static solution of these equations is very simple
\eqn\statsol{\phi_0= {1\over\pi}\sqrt{x^2-{1\over g}} }
where $|x|\ge \sqrt{{1\over g}}$. Note that $\phi_0$ actually
 makes the term inside the parenthesis in  Eq.\eqmotCOLL\ vanish.
That is, $\phi_0$ is a solution of the conventional field theory
equations  of motion Eq.\naive.

In the following we want the HD regime
of $x$ to possess this static solution. Hence we must take the HD limit
only for $|x|\ge \sqrt{{1\over g}}$.  Also it is important to recall that
   $\phi_0$  is the average eigenvalue density. Clearly,
for consistency $\phi_0>>1$. Note, however that $\phi_0$  decreases and
then vanishes as $|x|$ approaches $\sqrt{{1\over g} }$.
Therefore, the HD regime  of $x$ is  further restricted to
$|x|>> \sqrt{{1\over g}}$.

%
\vskip2in
\centerline{Figure 1. The potential and classical solution.}

What happens if one tries to ignore that restriction
 and tries to use the HD limit in the whole region $|x|\ge \sqrt{{1\over g}}$?
It comes back and shows itself in a different  and interesting way.
To explain the different disguises in which the same problem presents itself,
we need to mention a few more facts about collective field theory. Namely,
its perturbation expansion, and in particular the large order behaviour of
that perturbation series.

To obtain the perturbation expansion one expands  $\phi$ around the classical
solution $\phi_0$
\eqn\shft{\phi=\phi_0+ {1\over\sqrt{\pi}} \pa_x\z }
Substituting this into the Lagrangian \lager\ yields

\eqn\lquad{L_{\rm eff}=\int dx \left\{
 {1\over 2\pi \phi_0} {\dot\z}^2-{\pi\over 2}\phi_0 (\pa_x\z)^2-
{\sqrt{\pi}\over 6} (\pa_x\z)^3+\cdots\right\} }
To obtain a canonical kinetic term for the field $\z$ change coordinates to
the Liouville coordinate
\eqn\liou{\tau={1\over\pi}\int^x {dy\over\phi_0} }
$L_{\rm eff}$ then becomes
\eqn\threint{L_{\rm eff}=\int d\tau\left\{
 \half{\dot\z}^2-\half (\pa_\tau\z)^2-
{1\over \pi^{3\over 2}\phi_0^2(\tau)}
{1 \over 6} (\pa_\tau\z)^3+\cdots \right\} }
It follows from the cubic term that  ${1\over \pi^{3\over 2} \phi_0^2(\tau)}$
is to be identified as the coupling parameter
of collective field theory. However, ${1\over\phi_0^2(\tau)}$ becomes
large exactly where the classical solution $\phi_0$ becomes small, which is
where the HD expansion breaks down. The field theory becomes strongly
coupled or in other words the semi-classical expansion, the loop expansion,
breaks down.

What happens if this fact is ignored ? Perturbation series  takes its revenge
by growing too fast. In fact, the $G$'th order in perturbation
expansion grows like $(2G)!$.
There are a few diagrams that become large \antalb. The main contribution
comes from the region  (in $x$ space) where the coupling parameter
becomes large.
This is very similar to the phenomenon of renormalons
\ref\ren{G. Parisi, Phys. Lett. B76 (1978) 65.} in ordinary
non-asymptotically free quantum field theory\foot{ We thank M. Moshe for
drawing our attention to this analogy.}. There the coupling
parameter grows  (in momentum space),
and there  are a few diagrams that receive large contributions from
that region of integration in  momentum space.

The behavior of the perturbation series of collective field theory is
different than the  growth of perturbation series caused
by instantons in ordinary field theory in two ways. The first is in the rate
it grows ((2G)! vs. G!), and the other is the way it grows. In ordinary field
theories with instantons the number of diagrams grows, and not their
magnitude.

Note, however, that all these phenomena are related to an expansion around
a  ``bad" static classical solution $\phi_0$. Therefore, like the situation
in ordinary field theories with instantons, other classical solution may
actually be responsible for that particular behaviour of the perturbation
series.

\subsec{Low Density Limit and One Eigenvalue Instantons}

The low density  (LD) limit of collective field theory is defined as follows.
Consider  a region of  $x$, denoted $J$,  of length
$l(J)$. The number of eigenvalues in this region is $N(J)$.
Then take the the double scaling limit in such a way that
\eqn\ldl{ {N(J)\over l(J)}\rightarrow finite}
It is clear that in this region of $x$ there is a finite number of
eigenvalues. The collective field $\phi$ now has a finite number of degrees
of freedom  in that region
 and the classical equations of motion \eqmotcoll\ become simply
\eqn\eqMOTCOLL{ {\pa\over \pa x} \left(
\int\limits^x dy \pa_t{\int\limits^y\phidot\over\phi}
-{1\over 2}{\int\limits^x\phidot \int\limits^x\phidot \over\phi^2}
 -{\pi^2\over 2} \phi^2 -   \half ({1\over g}-x^2)\right){}_{|x=\l_i(t)} =0 }
where the index $i$ now runs over $i=,1,...,N(J)$.

In this paper, we are particularly concerned with LD regions containing a
single eigenvalue $\l^*(t)$. Using the relation
\eqn\coleig{\phi(x,t)=\d(x-\l^*(t))}
the equation of motion \eqMOTCOLL\ (including the contact terms) simply becomes
\eqn\spoon{ {d^2\l^*\over dt^2} =  \l^*(t)}
Note that this equation is identical to Eq.\motmat\ evaluated for
a single eigenvalue with $V(\l^*)=V(0)-\half (\l^*)^2$.
There are no non-trivial static solutions of Eq.\spoon.
The only reason that we can get a static solution
in  the HD region collective field theory is due to the interactions
between eigenvalues.
The  interaction energy balances the potential energy and the ``particles"
are at rest. The solutions that we are looking for are,
therefore,  time dependent.

The general solution of Eq.\spoon\ is given by
\eqn\solone{\l^*(t)=E^* cosh t+ F^* sinh t}
where $E^*$ and $F^*$ are real constants.  The energy of this solution
is
\eqn\energsol{E=V(0)-\half \left((E^*)^2-(F^*)^2\right)}
As discussed in the previous
subsection, we have restricted the HD region to $|x|\ge \sqrt{{1\over g}}$.
Therefore, in this paper, the LD region is in the complementary region
$-\sqrt{{1\over g}}\le x\le \sqrt{{1\over g}}$.

It is clear that almost all solutions \solone\ leave the LD region after a
finite time. The only exceptions are the two solutions corresponding to
$|E^*|=|F^*|$ which end up sitting on top of the potential
at time $t\rightarrow\infty$. These do not have an obvious physical interest.

So far we  have discussed classical solutions of the equations of motion in the
LD region in real Minkowski time. These
correspond to a finite number of eigenvalues  moving in real time.
More interesting are the classical solutions of the equations of
motion in the LD region in Euclidean time.
These are instantons that correspond to tunneling  of  eigenvalues
across the potential  barrier.
Going to Euclidean time  the
resulting equation of motion becomes
\eqn\ruland{ {\pa\over \pa x} \left(-
\int\limits^x dy \pa_\t{\int\limits^y\phidot\over\phi}
+{1\over 2}{\int\limits^x\phidot \int\limits^x\phidot \over\phi^2}
 -{\pi^2\over 2} \phi^2 -   \half ({1\over g}-x^2)\right){}_{|x=\l_i(\t)} =0 }

where $\t$ is Euclidean time.  Again we are particularly interested
in a LD region containing a
single eigenvalue $\l^*(\t)$. Using the relation
\eqn\coleigeuc{\phi(x,\t)=\d(x-\l^*(\t))}
the equation of motion \ruland\ (including the contact terms) simply becomes
\eqn\spooneuc{ {d^2\l^*\over d\t^2} =  -\l^*(\t)}
The general solution of Eq.\spooneuc\ is given by

\eqn\soloneeuc{\l^*(\t)=A^* cos \t+ B^* sin \t}

where $A^*$ and $B^*$ are real constants.
As in Minkowski space, the Euclidean  LD region is taken to be
$-\sqrt{{1\over g}}\le x\le \sqrt{{1\over g}}$.

As an example consider the collective field \coleigeuc\ corresponding to a
single eigenvalue $\l^*$ with the boundary conditions
$\l^*(\t_0)=\sqrt{{1\over g} }$ and ${\dot\l}^*(\t_0)=0$. It follows from
\soloneeuc\ that

\eqn\soloneeeuc{\l^*(\t) =\sqrt{{1\over g}} cos(\t-\t_0) }
The solution is shown in Figure 2.
\bigskip
%
\vskip2in
\centerline{Figure 2. A single eigenvalue $\l^*$ in Euclidean space.}
The corresponding collective field configuration is

\eqn\oneinst{
\phi_{inst}(x,\t)=\d\left(x-{1\over\sqrt{g}} cos(\t-\t_0)\right) }

This is an instanton that corresponds to tunneling of one eigenvalue
 across the barrier from
$x=\sqrt{{1\over g}}$ at $\t=\t_0$ to
$ x= -\sqrt{{1\over g}}$ at $\t=\t_0+\pi$. Note that the classical
solution in Eq.\oneinst\ is not a solution of the Euclidean continuation  of
of the unconstrained  field theory equations of motion Eq.\naive.

The action of the instanton $\phi_{inst}$
 can be computed from the Euclidean continuation of \lager. The result is
\eqn\actoneinst{S_{\rm eff}[\phi_{inst},\phidot_{inst}]=\int_0^{\pi} d\t
\int\limits_{+\sqrt{{1\over g} }}^{-\sqrt{{1\over g} } }dx
\d\left(x- {1\over\sqrt{ g}}
cos(\t)\right)\left\{ {1\over 2 g}\sin^2(\t) -{1\over 2 g}\cos^2(\t) +
{1\over g} \right\}= {\pi\over g} }
in agreement with the large order behavior of the perturbation series in $g$.
We want to stress that the action  of this solution
is not infinite as one might have thought.  The local
interaction  terms look infinite. For example, the $(\phi_{inst})^3$ term
gives a contribution proportional to $(\d(0))^2$. However, this contribution
gets canceled, due to the self-energy subtraction terms.
These terms also fix the normalization
of the field configuration  that corresponds to a single eigenvalue.
If we tried a configuration of the form
$\phi=A \d(x-\l^*(t))$ where $A\ne 1$ it would have  had
infinite energy.

Now let us see why we obtained an action of  ${1\over g}$
and not ${1\over g^2}$, as expected in ordinary field theory.
Collective field theory has two  ingredients that make this possible.
The first is the presence of scale breaking terms mentioned in
the introduction, i.e. the coupling parameter $g$ cannot be scaled out
from the  Lagrangian. The second and related ingredient is that the
solution is a  constrained solution  that depends on the  coupling parameter
in a way which is of  course different from what could be expected for a
solution of the $\phi$  equations of motion.

We can now say that we understand why the perturbation expansion of collective
field theory in the double scaling limit is growing as it grows. Besides
the static solution, there are time dependent Euclidean (constrained) solutions
that contribute to the path integral. The important configurations are
singular configurations, but have finite action. Perturbation theory is
smart enough and knows about these solutions. It lets us know of their
existence by growing accordingly.

\newsec{SPACE-TIME EFFECTIVE ACTION}

The collective field theory Lagrangian density, expression \lager, has
two notable deficiencies. First, the kinetic energy term is not
in canonical form. This means that we have not identified correctly
the  canonical field of the theory.
Second, and more important, the  coordinate $x$ appears in the
potential  energy and therefore Lorentz invariance seems to be
broken explicitly. In this section we remove both deficiencies.
The first, following \das, by  field and coordinate redefinitions.
The second, following
ref.\ref\ramyshanta{R. Brustein and S. De Alwis, Phys. Lett. B272 (1992) 285.},
 by enlarging the theory to include a new field. The non-trivial
vacuum expectation value of this  new field is responsible for the spontaneous
breaking of  Lorentz invariance.

\subsec{Canonical Collective Field }

As we demonstrated in section 3, the collective field theory divides
naturally into  three regions shown in Figure 3.
 Regions $I$ and $III$ admit the static solution $\phi_0$
given in Eq.\statsol. Region $II$, on the other hand, does not admit a static
solution. It does, however, possess Euclidean time dependent solutions.

The two high density regions $|x|>\sqrt{{1\over g}}  $,  are denoted
by I and III in Figure 3. The low density  region
$|x|\le\sqrt{{1\over g}}$, is denoted  II.

%
\vskip2.5in
\centerline{Figure 3.  High and low density regions}
 We begin the discussion in this section with region $I$.
The first step is to shift the double scaled collective field $\phi$ by the
classical static solution $\phi_0$ in Eq.\statsol.
\eqn\shfta{\phi=\phi_0+{1\over\sqrt{\pi}}\pa_x\z }
The resulting Lagrangian is $L= L_\z+L_0$ where
\eqn\lzeta{L_\z=\int dx \left\{
  \half {1\over \pi} {{\dot\z}^2 \over \phi_0+{1\over\sqrt{\pi}}\pa_x\z }
-\half\pi\phi_0 (\pa_x\z)^2-
{\sqrt{\pi}\over 6}(\pa_x\z)^3 \right\} }
and
\eqn\lzero{ L_0=\int dx \left\{-{\pi^2\over 6}(\phi_0)^3-
\half({1\over {g}}- x^2)\phi_0\right\} }
To obtain a canonical kinetic term for the field $\z$,  change coordinates to
the Liouville coordinate defined by

\eqn\liou{ \tau-\tau_0^I
={1\over\pi}\int\limits^x_{x_0} {dy\over\phi_0}
=\ln\left[x+\sqrt{x^2-{1\over g} }\ \right]
-\ln\left[x_0+\sqrt{x_0^2-{1\over  g}}\ \right]}

For simplicity, we take
$x_0=\sqrt{{1\over g} }$ and $\tau_0^I=\ln\sqrt{ {1\over g}} $. In this case

%
\eqn\coors{x= \sqrt{{1\over g}}\cosh(\tau-\ln\sqrt{{1\over g}}) }
and the range of  the two coordinates is
\eqn\rng{\eqalign{
\sqrt{{1\over {g}}}\le & \ x\ \le \infty \cr
\ln[\sqrt{{1\over {g}}}\ ] \le &\  \tau\ \le \infty
\cr   } }

The static solution in the new  coordinate becomes
\eqn\stattau{\phi_0(\tau)=
{1\over \pi \sqrt{g}}\sinh (\tau-\ln\sqrt{{1\over g}}) }

Other choices of $x_0,\tau_0^I$ in Eq.\liou\ would lead to an overall rescaling
of $\phi_0(\tau)$ and a rescaling of $g$ in Eq.\stattau.

Rewritten in terms of the Liouville coordinate the classical
Lagrangian, $L_\z$, is given by
\eqn\lzetau{L_\z  =\int d\tau \left\{
 \half{{\dot\z}^2 \over 1+{1\over \pi^{3\over 2}\phi_0^2(\tau)} \pa_\tau\z}
-\half (\pa_\tau\z)^2
-{1\over 6 }{1\over \pi^{3\over 2}\phi_0^2(\tau) } (\pa_\tau\z)^3 \right\}}
where $\phi_0$ is given by Eq.\stattau. Note that now the kinetic
term of $\z$ is indeed $\half\left( {\dot\z}^2- (\pa_\tau\z)^2\right)$
as it should be.

The pure $\phi_0$ Lagrangian , $L_0$, turns into
\eqn\lzertau{L_0=\int d\tau {\pi^3\over 3}\phi_0^4}
{}From the cubic interaction term in $L_\z$, it follows that
the coupling parameter of collective field theory is
\eqn\colcoup{{ 1\over \pi^{3\over 2}\phi_0^2(\tau)}=
4\sqrt{\pi} {e^{-2\tau}\over\left(1-{1\over {g}} e^{-2\tau} \right)^2} }

The coupling parameter vanishes as $\tau\rightarrow\infty$ and explodes at
$\tau=\ln[\sqrt{{1\over {g}}}\ ]$.

\subsec{Lorentz Invariance}

We now describe a field theory that reduces to  the collective field
theory  of region $I$ when the various fields obtain their expectation values.
The idea was discussed  for  $\mu=0$  in  \ramyshanta.
We limit ourselves here to flat target space, but $\mu\ne 0$.
 We note
that the $\z$ theory is not Lorentz invariant. Our interpretation is
that this is really a Lorentz invariant field theory of two fields,
$\z$ and $D$, expanded around the vacuum expectation values of the two fields.
The new field $D$ has a vacuum expectation value that breaks Lorentz
invariance, and that is the reason that the $\z$ theory alone is not
Lorentz invariant.

We also know that the true theory is defined for all $\z,D$
field configurations  and should  not be expressible  just as an expansion
around a particular solution. We therefore  look for a field theory which
has the appropriate solutions.

To find out the background independent field theory we have to identify
the expectation value of the $\z$ and $D$  fields first.
Motivated by the comparison between collective field theory and the
Polyakov description of the related string theory (see section 6)
we postulate that
\eqn\ohsol{
\eqalign{<G_{\mu\nu}> & =\eta_{\mu\nu}\cr <D> & = - 2\tau \cr
<\z> &={1\over g}  \cr} }
Here we added the expectation value of the metric as well. Our convention
is $\eta_{\mu\nu}=\pmatrix{1 & 0 \cr 0 & -1}$. Note that
the field  $D$ has the non-translation invariant vacuum expectation value.

We list the background independent form of the different quantities.

\eqn\stattawo{\pi^{3\over 2}\phi_0^2(\tau)\rightarrow
{1 \over 4\sqrt{\pi} } e^{-D}\left(1-{1\over {g}} e^{D} \right)^2  }
\eqn\covexp{ \pa_\tau \z \rightarrow \half \na D\cdot\na\z}
\eqn\anotther{{\dot\z}^2-(\pa_\tau \z)^2\rightarrow\na\z\cdot\na\z}
The Lorentz non-invariant quantities on the left are obtained from the Lorentz
invariant quantities on the right by letting $D=<D>$ and $\z=<\z>+\z'$.

Also, let
\eqn\covexpa{\int dt d\tau=\int d^2x}

We can now write the $\z$ action using the previous dictionary of
expressions. The result is

\eqn\acna{\eqalign{
 & {\cal S}_\z=  \int d^2x  \Biggl\{
\half{\na \z\cdot\na \z\over 1+ 2\sqrt{\pi}
{ e^{D}\over\left(1- {1\over {g}} e^{D} \right)^2 }\na \z\cdot\na D}
\cr &
- { \sqrt{\pi}\over 4 }{ e^{D}\over\left(1- {1\over {g}} e^{D} \right)^2 }
{(\na \z\cdot\na D)^3\over 1+2\sqrt{\pi}
{ e^{D}\over\left(1-{1\over {g}} e^{D} \right)^2 }\na \z\cdot\na D}
 - { \sqrt{\pi}\over 12}
{e^{D}\over\left(1-{1\over {g}} e^{D} \right)^2 }
(\na \z\cdot\na D)^3 \Biggr\} \cr } }
We also need the action for the pure $D$ sector. This action is not
supplied in such a clear way  by collective field theory, although
some hints are given in \lzero. We also know some of the lowest order terms
through other methods of calculation.

Recall the pure $\phi_0$ terms  given in Eq.\lzertau.
They can be  reexpressed using the dictionary of expressions
Eqs.\stattawo -\covexpa\ as
\eqn\lzerot{S_0=-
{1\over 384 \pi}\int d^2x e^{-2 D}
 \left[1-{1\over {g}} e^{D} \right]^4
\left[ {(\na D)^2-4} \right] }

Since space-time is flat in this case  we ignore  curvature terms in the
action. Higher derivative terms like $(\na D\cdot\na D)^2$ cannot be ruled
out at this point. In what follows we treat the Lagrangian \lzerot\ as
if it were the exact Lagrangian. Therefore,
the full space time action is
\eqn\monster{\eqalign{
 & {\cal S}=  \int d^2x\times  \cr & \Biggl\{
\half{\na \z\cdot\na \z\over 1+{ 2\sqrt{\pi} }
{ e^{D}\over\left(1- {1\over {g}} e^{D} \right)^2 }\na \z\cdot\na D}
- { \sqrt{\pi}\over 4 }{ e^{D}\over\left(1- {1\over {g}} e^{D} \right)^2 }
{(\na \z\cdot\na D)^3\over 1+{ 2 \sqrt{\pi} }
{ e^{D}\over\left(1-{1\over {g}} e^{D} \right)^2 }\na \z\cdot\na D}
\cr  -&  { \sqrt{\pi}\over 12}
{e^{D}\over\left(1-{1\over {g}} e^{D} \right)^2 }
(\na \z\cdot\na D)^3 -
{1\over  384 \pi} e^{-2 D}
 \left[1-{1\over {g}} e^{D} \right]^4
\left[ {(\na D)^2-4} \right]
\Biggr\} \cr }  }

Let us now discuss the equations of motion  derived from this action and check
that our construction is indeed consistent.
The equations of motion are complicated
 so we do not write them out explicitly.
However, it is straight forward to verify that indeed
\eqn\sssol{\eqalign{<G_{\mu\nu}> & =\eta_{\mu\nu}\cr <D> & = - 2\tau \cr
<\z> & ={1\over g} \cr} }
is an exact solution of these equations of motion.
There is also agreement with
the $\sigma$-model calculation  (see section 6.)
This is a non-trivial consistency check.

The solution \sssol\ is not a unique solution of the equations of motion.
The action \monster\ is Lorentz invariant. The solution \sssol\ breaks
that invariance and, therefore, there should be a  family of
solutions associated with the broken generators. Indeed the most general
solution for $D$ is (see \joe,\ramyshanta)
\eqn\dilsol{\eqalign{<D>= & a (t-{\bar t}) +
b (\tau -{\bar\tau})\cr  & b^2-a^2=4\cr } }

The solution \sssol\ corresponds to $a=0, b=-2,{\bar\tau}=0$.

Because the field $\z$ has only derivative couplings,
the action \monster\ is also invariant under  a $\z$ shift symmetry
\eqn\shiftsymm{ \z\rightarrow \z+ constant}
which is again reflected
in the fact that any $\z=constant$ is an allowed solution. We see now that the
original identification of the relation between the coordinate $\tau$ and
$x$ in Eq.\coors\ was not unique. However each definition corresponds to
a single and particular choice of $a,{\bar t},{\bar \tau}$ in Eq.\dilsol.

Note that the action \monster\ reduces to the one obtained in \ramyshanta\
in the limit $\mu\rightarrow 0$. To see that, replace $\phi_0$ in Eq.\stattau\
 by it's $\mu=0$ expression, $\phi_0^{\mu=0}={1\over 2\pi} e^{\tau}$.

There is a very important difference between  the theories with
$\mu=0$ and with $\mu\ne 0$. It is the appearance of what we
call ``scale breaking terms".
For  the  case $\mu=0$  the coupling
parameter of the effective field theory is expressed as a function of
$D$ alone.  When $\mu\ne 0$, this is not possible.
The obstruction comes from the fact that
the efective field theory coupling parameter, $\hbox{\Bigit g}$,
depends on both  ${g}={1\over\mu}$
and  $e^D$, and not just on $ e^{D}$. It is impossible to absorb $g$ into
a redefinition of $D$. As explained in the introduction,
this fact is closely related to the  behaviour of large order perturbation
 series and to appearance of finite action
configurations with action ${1\over g}$ as opposed to
${1\over g^2}$ .

\subsec{Extension to All  of Space Time}

At this point, we must recall that the effective Lagrangian \monster\
has been constructed only  for region $I$ of collective field theory. Hence,
the spatial parameter $\tau$ is restricted to satisfy
$\ln [\sqrt{ {1\over g}} ]\le\ \tau\ \le \infty$. It is essential that the
effective field theory be defined for all $\tau$ in the range
$-\infty\ \le \tau\ \le \infty$. It is possible to achieve this by
appropriate modification of the previous discussion in regions $III$ and $II$.
We find it easier, however,
to simply postulate that effective field theory \monster\ is valid
for all $\tau$, as it must be and then work  our way backward and show
that effective field theory  reduces to collective fields theory in
regions $III$ and $II$.

 In region III
the extension is straight forward. Since region $III$ is also a high density
region where the static solution  $\phi_0$ exists,
collective field theory Lagrangian
\lzeta\ and \lzero\  remain valid in that region. Similarly, the relationship
between $x$ and $\tau$ is almost identical to that in region $I$. However,
there are two trivial differences. Region $III$ runs from
$-\infty\ \le x \le \sqrt{ {1\over g}}$. Therefore Eq.\liou\ must be
modified in region $III$ to be
\eqn\liouIII{\tau-\tau_0^{III}=-{1\over\pi}\int\limits_{x_0}^x{dy\over\phi_0}}
where $x_0=-\sqrt{ {1\over g}}$. The second difference is that, having
chosen $\tau_0^I= \ln[ \sqrt{ {1\over g}}\ ]$ in region $I$,
we are no longer free to choose $\tau_0^{III}$ in region $III$.
It is determined by matching conditions
with region $II$. Solving Eq.\liouIII\ for $x$ we find

\eqn\exte{ x= -{1\over \sqrt{g}} \cosh(\tau-\tau_0^{III}) }

The range of the two coordinates is
\eqn\rngIII{\eqalign{
\infty \le & \ x\ \le  -\sqrt{{1\over {g}}} \cr
-\infty \le &\  \tau\ \le \tau_0^{III}  \cr   } }
The static solution in the new  coordinate becomes
\eqn\stattauIII{\phi_0(\tau)=
- {1\over \pi\sqrt{g}} \sinh(\tau-\tau_0^{III}) }
Collective field theory Lagrangian \lzetau\ and \lzertau\ remain
valid in region $III$, except that $\phi_0$ is given by
Eq.\stattauIII.
Now assume that effective
field theory \monster\ remains valid in region $III$.
It is straightforward to show that
\eqn\sssolIII{\eqalign{<G_{\mu\nu}> & =\eta_{\mu\nu}\cr
<D> & =2 \left(\tau-\left[\tau_0^{III}
-2\ln[ \sqrt{ {1\over g}}\ \right] \right)\cr
<\z> & ={1\over g} \cr} }
is an exact solution of the associated equations of motion.
Writting $D=<D>$ and $\z=<\z>+\z'$, and inserting this into
effective field theory \monster\ yields collective field theory
Lagrangian  \lzetau\ and \lzertau\ with
$\phi_0$ given by Eq.\stattauIII. Hence effective field theory \monster\
is indeed valid in region $III$.

Region II is more complicated. The static solution $\phi_0$ is not
defined there, and hence, Eq.\shfta\  must be modified to
\eqn\shftb{\phi=\pa_x\z}
It follows that $L=L_\z+L_0$, where the $\z$ Lagrangian is
\eqn\lzetaII{L_\z=\int dx \left\{ { {\dot\z}^2\over\pa_x\z }
-{\pi^2\over 6} (\pa_x\z)^3 -\half({1\over g}-x^2)\pa_x\z\right\} }
and $L_0=0$. Furthermore, it is
impossible to naively extend the relationship between
 $\tau$  and $x$ to  region II
because then  the coordinate $\tau$ would become imaginary.
Following the usual analytic continuation in matrix model
we define the relation between $x$ and $\tau$ to be what it would have been if
$\phi_0$ was the analytic continuation of the classical solution, i.e.
$\sqrt{{1\over g} -x^2} $.

The relation between $x$ and $\tau$ is therefore
\eqn\randal{\tau-\tau_0^{II}=
 -\int\limits_{x_0}^x {dy\over\sqrt{ {1\over g}-y^2} } }
where $x_0=\sqrt{{1\over g} }$.
Again $\tau_0^{II}$ must be determined by matching
conditions with region $I$. Solving Eq.\randal\ for $x$ we find
\eqn\exteexte{ x=
\sqrt{{1\over g}}  \cos{(\tau-\tau_0^{II} )} }
The range of the two coordinates is
\eqn\rngII{\eqalign{
 -\sqrt{{1\over {g}}}\le & \ x\ \le  \sqrt{{1\over {g}}} \cr
\tau_0^{II}-\pi \le &\  \tau\ \le \tau_0^{II} \cr   } }

At this point we can solve for $\tau_0^{II}$ and  $\tau_0^{III}$.
We demand that the three regions of $\tau$ given in Eqs.\rng, \rngIII\ and
\rngII\ continuously match onto one another at their common boundaries.
This implies that
\eqn\solmatch{\eqalign {
\tau_0^{II} & = \ln[\sqrt{{1\over {g}}}\ ]\cr
\tau_0^{III} & = \ln[\sqrt{{1\over {g}}}\ ] - \pi \cr }}
The coordinate $\tau$ is now defined continuously over the  whole real axis
$-\infty \le \tau \le\infty$ as required.

It is clear however that to any other choice of $\tau^{I}_0$ in Eq.\liou\
corresponds another set of matching conditions. The only difference is
that  regions $I,II,III$ will be displaced by the appropriate amount.
This is a reflection of the spontaneously broken translation invariance
of the field theory.

In terms of the coordinate $\tau$, the Lagrangian \lzetaII\ becomes

\eqn\lzetauII{L_\z=\int d\tau \left\{
{1\over {1\over\varphi_0^2}\pa_\tau\z } \half\left({ {\dot\z}^2}
-(\pa_\tau\z)^2\right)
-{\pi^2\over 6\varphi_0^2} (\pa_\tau\z)^3 \right\} }
where
\eqn\vpz{\varphi_0(\tau)= \sqrt{{1\over g}} \sin(\tau-\tau_0^{II})}

Comparing eq.\lzetauII\ against $L_\z$ in region $I$ and $III$, we see that
$L_\z$ in region $II$ is of a completely different structure. This difference
can be traced to  two sources. The first is that $\phi_0$ does not exist in
region $II$. The second is that the quantity $\varphi_0$, which takes
the place of $\phi_0$ in region $II$, is really an imaginary extension
of $\phi_0$ in the sense that $\varphi_0=-i\phi_0$. We expect, therefore,
that we will have trouble extracting the region $II$ collective field theory
from the effective field theory.
Let us try anyway. Assume the effective field theory \monster\ remains valid
in region $II$. Then it is straight forward to show that if we expand

\eqn\fconexp{\eqalign{D & = <D> \cr
            \z & =<\z>+i\sqrt{\pi}\z' \cr }}
where
\eqn\fconfcon{\eqalign{
<G_{\mu\nu}> & =-\eta_{\mu\nu} \cr
<D> & =2 i (\tau- \tau_0^{II})-2\ln[ \sqrt{ {1\over g}}\ ] \cr
<\z> & ={i \over 4 g \sqrt{\pi} }\left[
\sin (\tau-\tau_0^{II})- 2(\tau-\tau_0^{II})\right]\cr } }
 and go through the same
steps as before, we do indeed reproduce collective field theory \lzetaII.

The role reversal between $\tau$ and $t$ expressed in Eq.\fconfcon\
resembles what happens in certain analytically continued
coordinate regions of a two dimensional black hole.

The expected trouble  presents itself in Eqs.\fconexp\ and \fconfcon\ in two
ways. First, \fconfcon\ is not a solution of
the effective field theory equations of motion and, second,  $<D>$ and
$<\z>$  are imaginary. These are a direct consequence of the two sources
of trouble mentioned above. In order to cancel the $\phi_0$ dependence
in the collective field theory, which is implicit in the effective field theory
\monster, it is necessary for $<\z>$ to have the $\tau$ dependence shown in
Eq.\fconexp. However, these terms do not satisfy the equations of motion.
Furthermore,  in order to obtain $\varphi_0$ instead of $\phi_0$,
it is necessary for $<\z>$  and $<D>$ to be imaginary.
We conclude from this that, although it is possible to
derive the collective field theory \lzetauII\ from the effective field theory
\monster,  the relationship is strictly formal the effective field  theory
is not an efficient way of  describing the dynamics.
The real reason for all these difficulties is that collective field theory in
region $II$ is a low density theory of a finite number of eigenvalues.
We do not expect such theory to be described efficiently
by a continuous effective field  theory. If we insist on describing
the low density regions in terms of  the effective field theory then the scale
breaking terms, discussed previously, allow for the peculiar instanton action
${1\over g}$ (see next section.).

\newsec{STRINGY INSTANTONS}

In the previous section we showed that a Lorentz invariant effective field
theory can be associated with matrix models in the double scaling limit.
In this section, we discuss solutions of this effective field theory, both
in Minkowski and Euclidean space-time. We describe instantons,
that are associated with tunneling  between different solutions and
present their effects as new effective terms in $D$,$\z$ theory.

\subsec{Solutions in  Minkowski Space-Time}

The action for the Minkowski space-time effective field theory is given
in Eq.\monster. We assume that, a priori, this action is valid everywhere in
space and time. It is important to note that all interaction terms in
\monster\ are proportional to
\eqn\ftcc{ \hbox{\Bigit g} (D)=4\sqrt{\pi}
 { e^{D}\over\left(1- {1\over {g}} e^{D} \right)^2 } }
and, therefore ${\hbox{\Bigit g}} (D)$ is the effective  coupling parameter
of the theory. It is straight forward to derive the equations of motion
associated with \monster. However, these equations are complicated and,
for that reason, will not be written down explicitly here. Of more interest is
the general solution, given by
\eqn\sssolmina{\eqalign{<G_{\mu\nu}> & =\eta_{\mu\nu}\cr
<D> & =   a (t-{\bar t}) + b (\tau -{\bar\tau}) \cr
<\z> & ={1\over g}+c  \cr} }
where $a,b,c,{\bar t}$ and ${\bar\tau}$ are real parameters, $b^2-a^2=4$ and
$c,{\bar t},{\bar\tau}$ are arbitrary.

Of particular interest in this section are the static solutions $a=0$.
In this case
\eqn\sssolmin{\eqalign{<G_{\mu\nu}> & =\eta_{\mu\nu}\cr
<D> & =   \pm 2 (\tau -\bar{\tau}) \cr
<\z> & ={1\over g} +c\cr} }

First consider the solution where
\eqn\white{<D>=-2(\tau-{\bar\tau}) }
Then the effective coupling parameter becomes
\eqn\ftccm{ {\hbox{\Bigit g}}_-(\tau-{\bar\tau})=4\sqrt{\pi}
 { e^{-2(\tau-{\bar\tau})}\over
\left(1- {1\over {g}} e^{-2(\tau-{\bar\tau})} \right)^2 } }
Note that ${\hbox{\Bigit g}}_-$ is a function of $\tau-{\bar\tau}$ and,
hence, changes in its value for different points in space. Furthermore,
${\hbox{\Bigit g}}_-(\tau-{\bar\tau})\rightarrow\infty$
when $\tau=\tau_0^I$ where
\eqn\barnet{\tau_0^I={\bar\tau}+\ln\sqrt{{1\over g}} }
It is not hard to show that for $\tau <\tau_0^I$, the vacuum  can be
 described by a $<D>=+2(\tau-{\bar\tau}')$ solution where
${\bar\tau}'={\bar\tau}+2\ln\sqrt{{1\over g}}$. Such solutions
 will be described later. Therefore we restrict $\tau$ to satisfy
$\tau\ge\tau_0^I$. There is no loss in generality by setting ${\bar\tau}=0$.
Then $\tau_0^I=\ln\sqrt{{1\over g}}$ and the effective coupling
becomes
\eqn\ftccmz{ {\hbox{\Bigit g}}_- (\tau) =4\sqrt{\pi}
 { e^{-2\tau}\over\left(1- {1\over {g}} e^{-2\tau} \right)^2 } }
We plot this function in Figure 4.

\bigskip
%
\vskip2in
\centerline{Figure 4. Space dependent effective coupling
$\hbox{\Bigit g}_-(\tau)$.}
\medbreak
The physical interpretation of this vacuum state is the following.
For spatial points $\tau>>\ln\sqrt{{1\over g}}$ the effective
coupling is small and physics is well described by the effective
field theory \monster. However, as $\tau$ approaches $\ln\sqrt{{1\over g}}$
from the right, the coupling parameter blows up and a region
of strong coupling is encountered. Therefore, as
$\tau\rightarrow\ln\sqrt{{1\over g}}$ the effective field theory ceases
to adequately describe physics. The region to the left of the barrier,
$\tau<\ln\sqrt{{1\over g}}$, is terra incognita.
The effective field theory is not valid in this region. Perhaps new,
previously unknown dynamics applies there. More of this shortly.

Now consider the solution where

\eqn\mcmahn{<D>=2(\tau-{\bar\tau}'') }
Then the effective coupling parameter becomes
\eqn\ftccp{ {\hbox{\Bigit g}}_+(\tau-{\bar\tau}'')=4\sqrt{\pi}
 { e^{2(\tau-{\bar\tau}'')}\over
\left(1- {1\over {g}} e^{2(\tau-{\bar\tau}'')} \right)^2 } }
Note that ${\hbox{\Bigit g}}_+(\tau-{\bar\tau}'')
\rightarrow\infty$ when $\tau=\tau_0^{III}$
where
\eqn\archer{\tau_0^{III}={\bar\tau}''-\ln\sqrt{{1\over g}}}
For $\tau>\tau_0^{III}$, the vacuum can be described by a
$<D>=-2(\tau-{\bar\tau}''')$ solution. Since these solutions have
been discussed above, we restrict $\tau$ to satisfy $\tau\le\tau_0^{III}$.
For the present, we will set ${\bar\tau}''=2\ln\sqrt{{1\over g}}-\pi$.
It follows that $\tau_0^{III}=\ln\sqrt{{1\over g}}-\pi$ and the effective
coupling becomes
\eqn\ftccpz{ {\hbox{\Bigit g}}_+(\tau)=4\sqrt{\pi}
 { e^{2(\tau-2\ln\sqrt{{1\over g}}+\pi)}\over
\left(1- {1\over {g}} e^{2(\tau-2\ln\sqrt{{1\over g}}+\pi)} \right)^2 } }

We plot this function in Figure 5.
%

\vskip2in
\centerline{Figure 5. Space dependent effective coupling
${\hbox{\Bigit g}}_+(\tau)$.}
\medbreak
The physical interpretation of this vacuum state is identical to the physical
interpretation of the vacuum state above. However in this case,
the small coupling region where the effective field theory is
valid is  $\tau<<\ln\sqrt{{1\over g}}-\pi$, whereas the strong coupling region
is $\tau {<\atop\sim}\ln\sqrt{{1\over g}}-\pi$. The region to the right of
the barrier, $\tau > \ln\sqrt{{1\over g}}-\pi$ is terra incognita.

The solutions \white\ and \mcmahn\ represent two vacuum states of the
effective field theory. There is, however, another possible vacuum structure
which is the combination of these two solutions. That is, take
\eqn\ozie{<D>=-2\tau}
for $\tau\ge\tau_0^I=\ln\sqrt{{1\over g}}$, henceforth called region $I$,
and
\eqn\smith{<D>=-2(\tau-2 \ln\sqrt{{1\over g}}+\pi)}
for $\tau\le\tau_0^{III}=\ln\sqrt{{1\over g}}-\pi$,
henceforth called region $III$.
The effective coupling parameter in region $I$ is ${\hbox{\Bigit g}}_-$ given
in Eq.\ftccmz\ and in region $III$ is ${\hbox{\Bigit g}}_+$ given in
Eq.\ftccpz. The spatial interval
$\ln\sqrt{{1\over g}}-\pi\le \tau\le \ln\sqrt{{1\over g}}$
is called region $II$. We plot the effective coupling parameters in Figure 6.
%

\vskip2in
\centerline{Figure 6. Effective coupling parameter for a combined solution .}

The physical interpretation of this vacuum state is the following.
In regions $I$  and $III$,
away from the boundary points, the effective coupling parameter
is small and physics is described by the effective field theory \monster.
As $\tau$ approaches $\ln\sqrt{{1\over g}}$ from the right and as $\tau$
approaches $\ln\sqrt{{1\over g}}-\pi$ from the left, the coupling parameter
blows up and a region of strong coupling is encountered. Region $II$ is
terra incognita. Perhaps new, previously unknown, dynamics applies there.
We note in passing that we have chosen the width of region $II$ to be $\pi$.
This choice agrees with our choice of matrix model potential.
It is possible, of course, to make the width of region $II$  arbitrary by
changing the value of ${\bar\tau}''$ in region $III$. This arbitraryness
is discussed in the next section.

The three solutions just discussed are the only possible types of vacua of
the effective field theory. Of particular interest is the last solution,
shown graphically in Figure 6.  If all we knew was the effective
field theory, then we would have no interpretation of physics in region $II$.
However, we know more than the effective field theory. We know  matrix
 models, and that the effective field theory is the high density limit of
a double scaled matrix model. Comparing the vacuum of Figure 6 to the
 matrix model solution in section 5, we know exactly how to describe
physics in region II. Physics  in that region is not described by
 effective field theory \monster, but
rather by the low density collective field theory \lzetauII.

The vacuum of  of this low density theory is the state corresponding to
 the situation when there are no eigenvalues in the region
$\ln\sqrt{{1\over g}}\le\tau\le\ln\sqrt{{1\over g}}-\pi$.
In this case, the vacuum solution of the collective field theory in
region $II$ is clearly
\eqn\lenny{<\phi>=0 }
We want to emphasize that $\phi=0$ is a solution of the true collective field
theory equations of motion \motphi\ even though it is not a solution
of the naive equations of motion \naive.

The solution  in eq.\lenny\ matches continuously onto the vacua of regions
$I$ and $III$. To see this consider region $I$ and note that the vacuum
solution \ozie\ and \sssolmin\  in the $D$  and $\z$
variables is equivalent, using  Eq.\shfta,
to the collective field theory vacuum
\eqn\dykstra{<\phi>=\phi_0(\tau) }
where $\phi_0$ is given in Eq.\stattau. It follows that
\eqn\flyers{\phi_0(\ln\sqrt{{1\over g}})=0 }
which matches Eq.\lenny\ continuously at the boundary
$\tau_0^I=\ln\sqrt{{1\over g}}$. Similarly, in region $III$ the $D$, $\z$
variables vacuum solution \smith\  and  \sssolmin\ is
equivalent to the collective field theory vacuum
\eqn\phillies{<\phi>=\phi_0(\tau) }
where $\phi_0$ is given by Eq.\stattauIII. It follows that
\eqn\eagles{\phi_0(\ln\sqrt{{1\over g}}-\pi)=0 }
which matches Eq.\lenny\ continuously at the boundary
$\tau_0^I=\ln\sqrt{{1\over g}}-\pi$. Furthermore, note that the conjugate
momentum $\Pi_\phi$ of the static solution $\phi_0$ vanishes in both regions
$I$ and $III$. Similarly, the conjugate momentum vanishes in region $II$, since
there are no eigenvalues there. Hence, the momentum matches continuously
across the boundaries.

The complete vacuum solution
written  in terms of the collective field  $\phi$ for all of $\tau$ space,
is shown in Figure 7.
\medbreak
%
\ \ \vskip2.5in
\centerline{Figure 7. Complete vacuum  solution .}
Recall that  the position of this vacuum solution in $\tau$-space was
fixed by choosing $\bar\tau=0$, thereby making
$\tau_0^I=\ln\sqrt{{1\over g}}$. Furtheremore,
the width of region $II$ was chosen to be $\pi$ by letting
$\bar\tau''=\ln\sqrt{{1\over g}}-\pi$. Note that, by adjusting $\bar\tau''$ to
maintain a fixed width $\pi$, the entire vacuum solution can be translated
anywhere in $\tau$-space by varying the value of $\bar\tau$.

Finally, let us recall the discussion of single eigenvalue solutions of the
Minkowski space-time equations of motion in section 3.3. It follows from
Eqs.\solone\ and \energsol\ that any real solution satisfying the
boundary condition
$\l^*(t_0)=\sqrt{{1\over g}}$, must have initial velocity
${\dot\l}(t_0)>-\sqrt{{1\over g}}$ in order to overcome the potential barrier
and connect region $I$ with region $III$. In this case, however,
the conjugate momentum $\Pi_{\l^*}={\dot\l}^*$ is non-vanishing at the
boundary. Therefore, such single eigenvalue Minkowski solutions do not
continuously connect the zero momentum static solution $\phi_0$ of
regions $I$ and $III$.

\subsec{Solutions in Euclidean Space-Time}

The action for Euclidean space-time effective field theory is easily
obtained from Eq.\monster\ by analytic continuation of the time variable
$t$ to Euclidean time $\t$. The exact form of the Euclidean action is not
of importance in this section and therefore we do not write it down explicitly.
What is important, is that the effective
coupling parameter and the static solutions of the Euclidean equation
of motion  are still given by Eqs.\ftcc\ and \sssolmin\ respectively.
It follows that in Euclidean space, the vacuum structure given in Eqs.\ozie\
and \smith, and pictorially displayed in Figure 6, is also valid.
Region $II$ is now described by the analytic continuation of the low density
collective field theory \lzetaII\ to Euclidean space. Unlike the situation
 in Minkowski space, there is now a non-trivial excitation of one eigenvalue in
region $II$ that connects the vacua of region $I$ and region $III$. This single
eigenvalue excitation in Euclidean space was constructed in section 3.3, and
presented in terms of the collective field theory in eq.\oneinst.
Rewriting this solution in the $\tau$ coordinate, we find
\eqn\instburt{\phi_{inst}(\tau,\t)=
{1\over\sin(\t-\t_0) }
\d\left (\tau-[\ln\sqrt{{1\over g}}-(\t-\t_0)]\right) }
This is an instanton which corresponds to the tunneling of a single
eigenvalue across the barrier from $\tau=\ln\sqrt{ {1\over g}}$ at
$\t=\t_0$ to $\tau=\ln\sqrt{ {1\over g}}-\pi$ at $\t=\t_0+\pi$. Note that
from Eq.\soloneeuc\  it follows that the velocity of the eigenvalue
at either side of the barrier vanishes. Therefore,
the Euclidean conjugate momentum of the
instanton in region $II$, matches continuously at the boundaries with the
vanishing  conjugate momentum of the static  vacua $\phi_0$
in regions $I$ and $III$.
We represent this tunneling process in Figure 8.
\bigskip
%
\vskip2.5in
\centerline{Figure 8. Instanton connecting regions $I$ and $III$.}
\medbreak
It is of some interest to reexpress this instanton in terms of the field $\z$
using Eq.\shftb. In terms of $\z$ the instanton is given by
\eqn\INSTANTb{\z_{inst}(\tau,\t)=
\Theta
\left(\tau-[\ln{1\over\sqrt{g}}-(\t-\t_0)]\right) }

Note that this solution matches continuously onto the vacuum value
of $\z$, $<\z>={1\over g} +c_{I}$ and $\z={1\over g} +c_I-1$ in regions
$I$ and $III$ respectively for the appropriate value of $c_I$.

Therefore, the instanton field configuration is simply
 a kink moving in Euclidean time.
The position of the kink is where the argument of $\Theta$
in Eq.\INSTANTb\ vanishes.

We want to stress that this configuration is not a solution of the
$D$, $\z$ effective field theory.  We have not included explicitly the
terms in the field theory that correspond to self energy
subtraction (see Eq.\ham). However once they are taken into account,
the instanton configuration
has a finite action ${\pi\over g}$, as shown in Eq.\actoneinst.
Finally, we note that the initial tunneling time at which the instanton
starts its journey across the barrier, $\t_0$, is arbitrary.

\subsec{Effective Operators}

In this section we  integrate over the instantons and represent
their effects as effective terms in the $D$,$\z$ theory. Since $\z$ is
the light field we restrict our attention to $\z$ operators.
The effective operators are especially important. They provide the
bridge between the discrete, low density,
 sector of the theory and the continuous sector.
A full analysis is beyond the scope of the present paper. We will
content ourselves with a discussion of how the procedure of integrating
out instantons is implemented in this particular case and
 derive  the  dominant operator induced by instantons.

For a general discussion of the issue  of integrating out instantons
in field theory see
ref.\ref\svz{M. A. Shifman, A. I. Vainshtein and  V. I. Zhakharov,
Nucl. phys. B165 (1980) 45.}.
The integration over instantons is performed in the dilute gas
approximation. This approximation is justified for small $g$.
The first step
in integrating out the instantons was done  in the previous
section by constructing the single instanton configuration.
The second step  is to identify on how many parameters the solution depends.
 The number of parameters is usually the number of broken generators of the
full symmetry group of the theory.
The  parameters  become collective coordinates and are
integrated over.

The third step
is to integrate out the instantons in the semiclassical approximation
and present the result as a sum over local
operators constructed   from $\z$ and its derivatives.
These are then inserted back into an effective action. Thus the
effects of the instantons are  included.

We start by identifying  the parameters in the  stringy
instanton.  They are :
\bigskip
1.  The position of the instanton in Euclidean space $\bar\tau$, $\t_0$.

2. The orientation of the instanton  in Euclidean space $\alpha$.
\bigskip

The parameters $\bar\tau$ and $\t_0$ were  defined in Eq.\sssolmina\
and \instburt. The parameter $\a$ is  related to the  parameter
$a$ in the Euclidean space continuation of Eq.\sssolmina,   $a=2\sin\a$.
Changing $\a$ results in the rotation of the vacuum solution
in $\tau-\t$ space.

There are three zero modes corresponding to  the three broken generators
of the Euclidean group associated with $\bar\tau$, $\t_0$, $\a$.
These have to be integrated and produce a volume
factor
\eqn\vol{Vol\propto \int d\bar\tau d\t_0  d\a}

 When we discussed
the various types of vacua, the width of region $II$ was another parameter that
appeared to characterize the vacuum solution. However, it does not
correspond to a zero mode. In fact it corresponds to the constant mode
of the  scale factor of the metric $G_{\mu\nu}$,
which is a massive mode. The width of region $II$
is arbitrary but fixed. It was chosen to be $\pi$ to agree with previous
matrix model and collective field theory calculations.

The instanton solution also depends on  ${1\over g}$, the constant mode of the
field  $\z$.  It corresponds to the generator of the spontaneously broken
 shift symmetry $\z\rightarrow\z+const$.  Although this broken symmetry
does have a zero mode associated with it, we do not integrate over it. The
reason is that theories with different values of $g$ are really
different  theories. Their coupling parameters are different, and
therefore also physical amplitudes.

We now proceed to the third and final step. The dilute gas summation
over instantons induces  effective terms in the $D$,$\z$ Lagrangian.
The most general  action   induced by instantons is
\eqn\induced{ \D S=\int d\tau d\t \{\sum\limits_n C_n O_n(\tau,\t)\}  }
where $O_n$ are local operators built from  $D$ and $\z$ and
their derivatives.

The coefficients $C_n$ can be computed by expanding the action around the
instanton background. We find it convenient here to use the  effective  field
theory and its formal, but accurate, relationship to the collective
field  theory  in region $II$, given by Eq.\fconexp\ and Eq.\fconfcon.

Therefore we write
\eqn\braves{ \z= <\z>+i\sqrt\pi( \z_{inst}+\z')}
Then

\eqn\expan{ \int [d\z'] e^{-S({\bar\z}_{inst}+i\sqrt{\pi}\z',<D>)}=
\int [d\z'] e^{-S_0 ({\bar\z}_{inst}, <D>)+
\d S({\bar\z}_{inst},<D>, \z')} }
where ${\bar\z}_{inst}=<\z>+i\sqrt\pi \z_{inst} $.
Both $<D>$ and $<\z>$ are defined in Eq.\fconfcon,
and $\z_{inst}$ is defined in Eq.\INSTANTb

The remaining integrals  in Eq.\expan\ are computed in the
semiclassical approximation, i.e. expanding $\d S$ to quadratic terms in
$\z'$ only and performing the resulting Gaussian integral.
To extract particular $C_n$'s one
has to compute appropriate expectation values and compare them to
the expectation values in the trivial vacuum.
In adapting this formalizm to our theory
we have to take special care because the instanton is not a
solution of the naive $D$,$\z$  equations of motion.

All the coefficients $C_n$ are proportional to the universal factor of the
exponent of the instanton action  and the remaining
factor depends on the  particular operator that is considered.
 Since the ``size" of
the instanton is $\sqrt{{g}}$ (recall Eq.\dsl),
the dimension of the operator
determines the $ g$ dependence of $C_n$.
\eqn\expand{ C_n= \hbox{\it \~C}_n g^{d(n)} e^{-{\pi\over g} } }
where
\eqn\dimdum{d(n)=[{\rm dimension}(O_n)]^{\half}-1  }
and $\hbox{\it \~C}_n$ is a numerical coefficient.  The coefficient
$\hbox{\it \~C}_n$ is not expected to be particularly large or
particularly  small.
For example, the operator $\na\z\cdot\na\z$ has  naive
mass dimension 2, and therefore it's coefficient is proportional to $ g^0$.
The unit operator has dimension 0, and therefore $C_0\propto {1\over g}$.

We are interested in large ${1\over g}$ that corresponds to small $g$.
In that case the dominant and most interesting operator  is the unit operator.
All other operators are suppressed by powers of $ g$.
The coefficient  of the unit operator is given by
\eqn\cosmos{ C_0= \hbox{\it \~C}_0{\scriptstyle {1\over g}} e^{-{\pi\over g}}}

The numerical coefficient $\hbox{\it \~C}_0$ is given by
\eqn\czero{ \hbox{\it \~C}_0=\lim_{g\rightarrow 0} g
{\int [d\z'] e^{\d S_2(\z_{inst},D_{inst}, \z')}
\over
\int [d\z'] e^{-S_2 (\z')}  } }

where $\d S_2$ and $S_2$ are the quadratic actions around the instanton and
trivial vacuum respectively.

The result in Eq.\cosmos\  was obtained in the background of a constant
field $<\z>={1\over g}$. Lorentz
invariance then dictates that at least for slowly varying fields the effective
operator depends on the full field $\z$ and not just its constant mode
${1\over g}$.  Therefore the final result for the  induced operator is
\eqn\effop{ \D L_0=\hbox{\it \~C}_0\z e^{-\pi\z}}

This operator breaks the  $\z$ shift symmetry. It induces a  runaway
non-perturbative potential for the field $\z$.

\newsec{STRING THEORY}

In the previous sections we discussed  the  space-time effective theory
associated  with  matrix models and collective field theory. We  now
 review the connection between the world sheet
 description  of $1+1$  dimensional  string theory and matrix models
and collective field theory. We compare solutions of  our
 effective space-time theory to  solutions of the  $\beta$-function equations.
We also discuss the stringy instantons of
subsection (5.2)  from  the  world  sheet point  of view.

\subsec{String Theory and Liouville  Theory}

The class  of $1+1$ dimensional string theories that  we are  interested  in
is  described  by the  following two dimensional $\sigma$-model
\ref\eli{S. Elitzur, A. Forge and E. Rabinovici,
Nucl. Phys. B359 (1991) 581.},\ref\tsey{A. Tseytlin,
Phys. Lett. B264 (1991) 311.},
\eqn\sigac{I={1\over 4\pi}\int d^2z
\sqrt{\hat g}\left\{\hat g^{\a\b}G_{\mu\nu}\pa_{\a}X^{\mu}
\pa_{\b}X^{\nu} +\hat R D(X)+2 T(X)\right\}}
where $\hat g_{\a\b}$ is the fixed world sheet metric  with Euclidean
signature and  $\hat R$ is the corresponding Ricci scalar.
The sigma model  field $X_\mu$  stands for  two  scalar
fields, $X_0 (z)$, and $X_1(z)$.
The   field $G_{\mu\nu}(X)$ is the target space metric,  assumed here  to  have
Euclidean signature, $D(X)$ is the dilaton,  and $T(X)$ is the tachyon.
The names of the fields are a little bit misleading.
They originate from the  form of the world sheet couplings of
these fields. In 26  dimensional Minkowski space  critical string
theories, the  tachyon is indeed tachyonic and the  metric and dilaton
correspond to massless fields. However, as we  see  shortly,  in $1+1$
dimensions,  the physical tachyon is  really massless and the metric
and dilaton  correspond to massive non-propagating  fields.

Consistent  string backgrounds  are described by conformal field  theories.
The conditions for  conformal invariance are determined in general  by
the equations $\beta=0$, where  $\beta$ is the beta-function of the theory.
Of course, the $\beta$-function equations can only be computed in some
perturbative scheme. The lowest order equations for the theory described
by Eq.\sigac\  are
\eqn\bz{\eqalign{    R_{\mu\nu}+2\nabla_\mu\nabla_\nu D & =  0 \cr
                  -\half \nabla^2 D+ (\nabla D)^2 + 4 & =  0 \cr
                  -\nabla^2 T+2 \nabla D\cdot\nabla T-4 T & =  0 \cr }}

We can compare Eqs. \bz\ to the lowest order equations  of motion
derived from \monster. From this comparison we deduce that, to this order,
the field $D$ appearing in \monster\ is the same as the dilaton $D$
in \sigac\ and that the field $\z$ in \monster\ is related to the tachyon
as follows
\eqn\compzt{ \z\propto   T  e^{-D}}

This result is rather remarkable. It says that the equations of motion
of the background fields for the class of string theories specified by
the action \sigac\ are, to lowest order, identical to the equations
of motion of the effective field theory extension of matrix models. Can we
extend this identification beyond lowest order? To do this, let us consider
the solutions of the equations of motion \bz. These solutions
 can be classified  into two families. The first family
 consists of solutions  with  constant metric
$G_{\mu\nu}$ , and therefore vanishing curvature. The second family
 consists of  curved space solutions.
We  do not  discuss  the second family of curved space solutions in this paper.

The flat space solutions are
\eqn\solbz{\eqalign{    G_{\mu\nu} & =  \d_{\mu\nu} \cr
                       D & = a(X_0-\bar X_0)+ b(X_1-\bar X_1)   \cr
                       T   & = m e^D  \cr }}
where $a^2+b^2=4$. Of particular interest is the static  solution
\eqn\statsolbz{\eqalign{G_{\mu\nu} & =\ \d_{\mu\nu}\cr
D\  & =-2 X_1\cr T\ &= m e^{-2 X_1} \cr} }

By substituting the static  solution of Eq.\statsolbz\ into the sigma model
\sigac\ and writing $X_1=\varphi$, the following  world sheet
conformal field  theory is  obtained
\eqn\lcft{ I=
{1\over 4\pi}\int d^2z\sqrt{\hat g}\left\{\hat g^{\a\b} \pa_{\a}X_0\pa_{\b}X_0
+ \hat g^{\a\b} \pa_{\a}\varphi\pa_{\b}\varphi-2 \hat R \varphi
+2 m e^{-2\varphi} \right\}}
which can be identified as the Liouville conformal field theory with $c_m=1$
matter. To obtain a Minkowski signature string theory one has to analytically
continue $X_0\rightarrow i\hbox{\Bigit t}\ \!(z)$, which then becomes
the time variable of target space. The field $\varphi$ corresponds to
the spatial dimension of target space.

Liouville  theory \lcft\  was  compared  to matrix  models and to  collective
field theory  by a number  of  authors. Specifically, these authors
studied the complete theory, going beyond the lowest order.
 The  conclusion is that they
describe of the same theory. Evidence to this effect was obtained
on many  levels e.g., see \igor,\grig, and
\ref\fraku{P. Di Francesco and D. Kutasov, Nucl. Phys. B375 (1992), 119.}.
For a review and more comprehensive list of references see ref.\antalc.
In particular, the  relation between $\z$  and $T$ in Eq.\compzt\ as well
as the linear relation between $\varphi$ and $\tau$ have been well
documented.

More important, from our point of view, is that, by computing scattering
amplitudes of fluctuations around the Liouville vacuum,
one can determine the equations of the original string backgrounds $D$ and $T$
beyond the lowest order. When these are compared to the full equations of
motion derived from Eq.\monster, one finds that they are identical, as long as
the two $D$ fields are identified, relation \compzt\ holds and the parameter
 $m$ in \compzt\ is chosen to be
\eqn\compmg{m={1\over g} }

We conclude, therefore, that
\bigskip
\centerline{ The string theories associated with the world
sheet action \sigac\  have the same  }
\centerline{
equations of motion and effective action for
their background fields as does }
\centerline{
the effective field theory for the matrix model
given in Eq.\monster. }
\bigskip
 Furthermore, even the low density region of the matrix model discussed
earlier is  expected to to describe physical
aspects of these string theories, such as their non-perturbative behaviour.
It follows that the discussion of vacua, single eigenvalue instantons and
induced operators given in the previous section is, in fact, applicable
to the string theories associated with \sigac.

\subsec{Minkowski  and Euclidean Space-Time Backgrounds}

Based on our discussion  of solutions  of space-time effective theory
of section  5 we can now  discuss  general flat space solutions of
 the non-linear  extension of Eqs.\bz.
Since they  are  the same  equations they  have the same solutions
as explained before. Each  solution corresponds  to a different  conformal
field theory. The world sheet description of the different possible conformal
backgrounds were discussed in
ref.\ref\\polch{D. Minic, J. Polchinski and Z. Yang,
Nucl.Phys. B369 (1992) 324.}. The different backgrounds fall
into three classes. All of them can be obtained from the conformal field theory
 \lcft\ by a change of coordinates.
The first class corresponds to the coordinate change
\eqn\chcoi{\eqalign{ X'_0 & =\g(X_0-  v \vphi) \cr
                     \vphi' &= \g(\vphi- v t) \cr }}
where $\g={1\over\sqrt{1-v^2}}$. This change of coordinates transforms \lcft\
into a theory with $c_m<1$.
The second class corresponds to the coordinate change as well as the analytic
continuation $X_0\rightarrow i\hbox{\Bigit t}$
\eqn\chcoii{\eqalign{ \hbox{\Bigit t}' & =\g(\hbox{\Bigit t}+ v \vphi) \cr
                     \vphi' &= \g(\vphi- v \hbox{\Bigit t}) \cr }}
 This change of coordinates transforms \lcft\
into a theory with $1<c_m<25$.
The third class of conformal field theories is obtained by interchanging
the role of $\varphi$ and $X_0$ in \lcft.

As we can see from the factor $\g$ in  transformation rules
\chcoi,\chcoii, Lorentz invariance creeps into there. It is impossible to
understand  that from the world sheet point of view. However,
from the point of view of effective field theory this is just a reflection of
it's Lorentz invariance. Every class of theories corresponds to
a different solution for the dilaton field. The first class and
half  of the third class corresponds to  Euclidean solutions as described in
section (5.2), the second class  and the remaining half of the third class
corresponds to Minkowski solutions described in section (5.1).

The coupling parameter of the string theory defined by the $\s$-model
\sigac\ is  $e^{D}$, and therefore, for  the static solution \statsolbz\
 is position dependent.
\eqn\efcp{{g_{\rm st}}(\varphi)=e^{-2\varphi }}
In general $g_{\rm st}$ is position and time dependent.
Note that the string coupling parameter in Eq.\efcp\ and effective field
theory coupling parameter in Eq.\ftccm\ are different. In fact, the string
coupling parameter remains finite for finite spatial coordinate,
while the effective field theory coupling parameters blows up at
the boundary of region $I$.

In most string theories discussed so far, it was believed that  the string
 coupling parameter and the coupling parameter  in the effective field theory
are the same. There is a simple world sheet  argument, based on the universal
form of dilaton interactions,  that actually  ``proves" that.
 However, in this particular  example it fails! Another parameter $g$
appears, and enters into the expression for the coupling parameter.

The stringy instantons of section (5.3) can now be partially described in
world sheet language. The two high density regions $I$ and $III$ can each
be described by a separate Liouville theory. The boundary of each region
where the effective coupling parameter becomes strong is known as the
``Liouville wall''. Region $II$ in between region $I$ and $III$ cannot
be described in terms of the  world sheet theory. It would correspond to
an infinitely strong  coupling parameter. The signal for  the break down
of the world sheet description in this case is the `` wall".
The only  hint that one can obtain from the world sheet theory,
 as to what  happens  behind the wall, is given by the  growth of
large  order amplitudes of tachyons.
Inside region $II$, behind the ``Liouville wall'',
new  dynamics unaccessible  to the  Polyakov  description
of string theory takes over.  It is described  by a version of string
theory  capable of describing discrete physics,
the matrix model and collective field theory. It allows for quantum mechanical
amplitudes for tunneling processes where part of space enters the wall
and emerges on the other side into a copy of the world that it left.

\newsec{ CONCLUSION and OUTLOOK}

In this section we  list some of the  directions in which the results
obtained in this paper can be extended.

An obvious extension, which we do not foresee any problems in performing
 and  is currently under investigation
\ref\ramburt{R. Brustein, M. Faux and B. Ovrut, Penn Preprint, To Appear.},
is to consider the supersymmetric extension of collective field theory
and matrix models. The supersymmetric collective field theory
is discussed in ref.
\ref\antald{A. Jevicki and J. P. Rodrigues, Phys. Lett. B268 (1991) 53. } and
the supersymmetric eigenvalue theory is discussed in ref.\dabh,
where it is shown to be equivalent to the  Marinari-Parisi model.
In ref.\ref\tond{J. P. Rodrigues and J. van Tonder, Witwatersrand preprint,
CNLS-92-02 (1992).}
all these theories are  shown to be  equivalent to each other. Therefore
there is a unique supersymmetric extension of matrix models and collective
field theory.

This supersymmetric theory does have one eigenvalue instantons. Their
action is again ${1\over g}$. In ref.\dabh\ it is shown that these instantons
break supersymmetry with the expected  strength of $e^{-{1\over g}}$.
It should be possible to write down a supersymmetric
space-time effective action analogous to Eq.\monster, identify
it's classical Euclidean solutions and obtain the corresponding
stringy instantons as well as the effective operators that they induce.
We expect that these operators break supersymmetry non-preturbatively.

The most interesting and crucial step in finding out whether or not
our stringy instantons, or some of their higher dimensional relatives,
 play an important role in string theory is to find
out their effects in 4-dimensional string theories. Since matrix
models lose most of their power in 4 dimensions, it is unlikely, but not
impossible, that a direct application of the same techniques would be
useful to study that question. We can however look for space-time solutions
that have the same general features of the solutions described in section 5.
The most important of those features is that the space (or space-time)
dependent  effective coupling parameter  is blowing up at a finite point
in space.

There is in fact a  class of
4-dimensional string  solutions that have this property.
They are described in refs.\ref\meir{R. Myers,
Phys. Lett. B199 (1987) 371.}--\nref\ab{I. Antoniadis, C. Bachas, J. Ellis
and D. Nanopoulos, Phys. Lett. B211 (1988) 393.}\ref\shanjoe{S. De Alwis,
J. Polchinski and R. Schimmrigk, Phys. Lett. B218 (1988) 449.}.
The main features of these cosmological solutions is that
they possess a space dependent effective coupling parameter that becomes
infinite at a finite space-time point.  The supersymmetric counterparts
of these solutions were constructed in \shanjoe.

It is tempting to conjecture that stringy instantons similar to our
stringy instantons connect different regions of space-time and that they
induce non-perturbative operators of the type discussed in section 5.
In that case these operators are expected to be proportional to
the universal factor
$$e^{-\sqrt{S}}$$
Here $S$ is a complex field
that naturally appears in the effective low energy supergravity  field
theory obtained from superstring theory. The dilaton is related to
the real part of $S$.
$$
<Re S>\sim {1\over g^2}
$$
Note that the non-perturbative effects considered previously in the
literature induced operators that were proportional to the  universal factor
$$e^{-S}$$

Since the coupling parameter $g$ is expected to be small, the difference
between these two universal factors is quite big.
This may have important phenomenological consequences.

\bigskip
{ACKNOWLEDGEMENT}:
It is a pleasure to thank Lee Brekke, Joanne Cohn, Rick Davis,
Antal Jevicki and especially  Shanta De Alwis for useful discussions.

This work was supported in part by the Department of Energy under
contract No. DOE-AC02-76-ERO-3071.


\listrefs
\bye